\documentclass[journal]{IEEEtran}

\usepackage{xcolor,soul,framed} 

\colorlet{shadecolor}{yellow}
\usepackage[pdftex]{graphicx}
\graphicspath{{../pdf/}{../jpeg/}}
\DeclareGraphicsExtensions{.pdf,.jpeg,.png}

\usepackage[cmex10]{amsmath}
\usepackage{array}
\usepackage{mdwmath}
\usepackage{mdwtab}
\usepackage{eqparbox}
\usepackage{url}
\usepackage[utf8]{inputenc}
\usepackage[english]{babel}
\usepackage{cite}
\usepackage{url}

\usepackage{breakurl}
\usepackage[breaklinks]{hyperref}

\newcommand{\revi}[1]{\textcolor{black}{#1}}

\begin{document}
\bstctlcite{IEEEexample:BSTcontrol}
    \title{Twenty-one key factors to choose an IoT platform: Theoretical framework and its applications}
  \author{Mehar~Ullah,~\IEEEmembership{Member,~IEEE,}
      Pedro~Juliano Nardelli,~\IEEEmembership{Senior Member,~IEEE,}
      Annika Wolff, Kari Smolander\thanks{The authors are with LUT University, Finland. This paper is partly supported by by Academy of Finland via: (a) ee-IoT  n.319009, (b) FIREMAN consortium CHIST-ERA/n.326270, and (c) EnergyNet  Fellowship n.321265/n.328869. A preliminary of version of this paper appeared in \cite{Ullah2019}.}}


\maketitle

\begin{abstract}
Internet of Things (IoT) refers to the interconnection of physical objects via the Internet. It utilises complex back-end systems which need different capabilities depending on the requirements of the system. IoT has already been used in various applications, such as agriculture, smart home, health, automobiles, and smart grids. There are many IoT platforms, each of them capable of providing specific services for such applications. Finding the best match between application and platform is, however, a hard task as it can difficult to understand the implications of small differences between platforms. This paper builds on previous work that has identified twenty-one important factors of an IoT platform, which were verified by Delphi method. We demonstrate here how these factors can be used to discriminate between five well known IoT platforms, which are arbitrarily chosen based on their market share. These results illustrate how the proposed approach provides an objective methodology that  can be used to select the most suitable IoT platform for different business applications based on their particular requirements.
\end{abstract}


%
\IEEEpeerreviewmaketitle


\vspace{-0.5cm}
\section{Introduction}

The Internet of Things (IoT) concept was first coined by Keven Ashton in 1999 during a presentation to Proctor and Gamble and later referenced by him in the MIT Auto-ID Center \cite{Greer2019}. IoT is one of the fastest growing technologies that is gaining momentum in the various domains like transportation, healthcare, industrial automation, education sector etc. The main idea of IoT is to connect the physical world with the digital world\cite{Al-Fuqaha2015}. The foundation technology for IoT is RFID technology, which is used to identify, track and monitor any object with RFID tags and allow microchips to transmit the identification information to a reader through wireless communication \cite{Jia2012}. Nowadays, IoT applications have already moved further away than just simple RFIDs, incorporating different sources of data collection from sensors.
This data stream needs to be moved somewhere where this (big) data can be processed using, for example, machine learning techniques.
This place is what we call an \textbf{IoT platform}.

For companies to run their specific IoT applications, an IoT platform is then needed. The IoT platform provides important services and features to applications: endpoint management, connectivity and network management, analysis and processing, data management, application development, security, event processing, monitoring, access control and interfacing \cite{Mineraud2016}.
From 2015 onward, there has been a rapid growth in IoT technologies so that the number of connected devices and platforms have steadily increased.
For example, there were 260 IoT platforms in 2015, which increased to 360 in 2016 and 450 in 2017 \cite{zana2017}. Due to the technological improvements, new IoT devices emerged and the requirements of the IoT applications and platforms changed \cite{Hejazi2018}.
Such a technological change creates many challenges for businesses, governments and companies, which have little experience about the infrastructure of IoT and IoT platforms. Selecting a suitable IoT platform among all existing options is a tricky task since this decision needs to incorporate not only the current needs but also the potential future ones \cite{Guth2017}. 

\revi{There are hundreds of IoT platforms in the market, most with similar functionality with differences related to their implementation and underlying technologies \cite{Guth2017}.
Our aim with this research is to first highlight the key building blocks of IoT for the understanding of functionality and significance of IoT, identify and verify the key factors of an IoT platform. With the key factors in hand, we could then propose an objective and general methodology to compare the different service providers. 
In this case, our study develops a theoretical framework that will support companies in selecting a suitable IoT platform for their business needs.
To carry out this study, we follow three steps: (i) data collection, (ii) data verification and characterization, and (iii) application of the proposed framework.
These are the questions that were used to guide our research: (1) What is IoT as well as its building blocks? (2) What are the important factors of an IoT platform? (3) What factors should be considered for selecting an appropriate IoT platform for specific organizations?}

This present journal paper covers and extends  our preliminary work \cite{Ullah2019} in which we have identified twenty-one IoT platform factors from the literature and verified those factors using the Delphi method. 
This contribution builds upon \cite{Ullah2019}  aiming at developing a theoretical framework by comparing the twenty-one key IoT platform factors with the features provided by the five popular IoT platforms.
Our goal is to provide an objective while general methodology that different organizations can apply when selecting the most suitable platform based on their particular needs.
In other words, this paper will support such organizations to carry out a detailed analysis of their own requirements and understand the key IoT platform factors in order to find the best match.

The rest of the paper is organized as follows. In Section II we review the IoT building blocks. In Section III we identify the key factors of an IoT platform. \revi{Section IV introduces the research method employed here, including the results of the Delphi study. Section V focuses on the comparison of the main IoT platforms.} Finally, Section VI concludes this paper indicating potential future research paths.

\section {IoT Building Blocks}

\begin{figure*}
  \begin{center}
  \includegraphics[width=2\columnwidth]{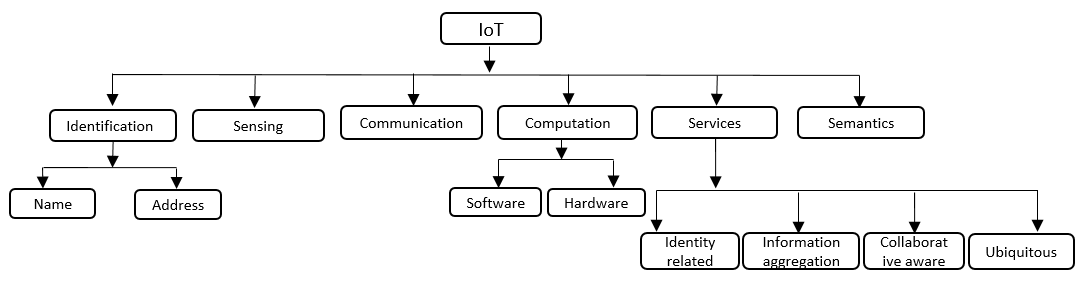}\\
  \caption{IoT building blocks, following the structure proposed by \cite{Okoli2011}.}\label{IoT-Building Blocks}
  \end{center}
  \vspace{-0.7cm}
\end{figure*}

To understand the functionality and significance of IoT, it is essential to understand its building blocks; they are the components of IoT, which work together to deliver its functionality. There are six IoT building blocks that work together and provide functionality  \cite{Al-Fuqaha2015}, as shown in Fig. \ref{IoT-Building Blocks}. 
In the following, we will explain each of them in more details.

\textit{Identification block:} The identification method is used to identify devices in the network. Devices are identified with the Object ID, which is the name of the device, and the object address, which provides the address of the device in the communication network \cite{Koshizuka2010}. The main addressing methods of IoT objects are IPv6 and IPv4 \cite{Al-Fuqaha2015}.   

\textit{Sensing block:} Sensors are used for collecting the data of objects/environment in the communication network and sending the collected data to the destination database or to the cloud. The data collected is analyzed in the  cloud. Actuators, i.e. hardware mechanical devices such as switches, are also used in IoT platforms and operate in the opposite way to a sensor \cite{Al-Fuqaha2015},\cite{Song2013},\cite{Mandal2016}.

\textit{Communication block: } It contains many heterogeneous objects that exchange data and various services with each other and with the  platform. The communication block contains IoT communication protocols like MQTT and CoAP that are used to connect different objects to IoT and to send data from those connected objects to the management system. The sensors and other devices are connected to the Internet by communication technologies like ZigBee, NFC, UWB, Wi-Fi, SigFox, and BLE \cite{Hejazi2018}, \cite{Al-Fuqaha2015}.

\textit{Computation block: }The computation block consists of two parts, hardware and software. Many hardware platforms have been built to run IoT applications, for example, Intel Galileo, Raspberry PI, Gadgeteer, UDOO, and Arduino. Similarly, there are many software platforms that are used to perform the functionalities of IoT. The main software platform is the operating system that runs throughout almost the whole activation time of the device. The cloud platform is also a computational component of the IoT; it enables small objects to send data to the cloud, it facilitates big data processing in real time and helps the end user to obtain knowledge extracted from the big data \cite{Hejazi2018},\cite{Al-Fuqaha2015}.

\textit{Services block:} IoT services aid IoT application developers by providing a starting point for development. When developers know the services available, they mainly focus on building the application rather than designing the service and architecture for supporting the IoT  application.  IoT   services  are divided into four categories. Identity related services can be divided into two categories, active and passive. Services that broadcast information and have a constant power or take power from the battery are active identity related services. Active identity related services can transmit or send information to another device. Passive identity related services have no power source and need some external device or mechanism to transmit its identity. Passive identity related services can only read information from devices. Information aggregation services refer to the actions of collecting data from sensors, processing that data, and transferring it to the IoT application for processing. Collaborative aware services use the data provided by the information aggregation services to make decisions and react accordingly.  Ubiquitous services  provide collaborative aware services anytime to anyone who needs it anywhere \cite{Song2013}, \cite{Gigli2011},\cite{Mohammed2015}.

\textit{Semantic Block:} IoT provides different services, for which it needs knowledge, and in order to get that knowledge in an effective way, IoT uses different machines. Knowledge extraction can include finding and using resources, modeling information, and recognizing and analyzing data to reach some decision and provide the correct service. So, it can be claimed that the semantic block is the brain of the IoT \cite{Hejazi2018},\cite{Al-Fuqaha2015},\cite{Song2013}.

\vspace{-1ex}
\section{Key Factors of an IoT platform}

An IoT platform is the main part of an IoT solution. There are hundreds of IoT platform vendors in the market, and finding and selecting a suitable IoT platform that is reliable and scalable is difficult. However, consideration of some key factors prior to making a platform selection decision can enable companies to find and select an appropriate IoT platform for their business. The platform requirements are context-specific and it is not necessary that a platform include all the factors discussed below, but can have a maximum. These factors were identified from literature by studying various IoT platforms \cite{amazon2019,Azure2019,things2019,google2019,IBMwat2019},  articles \cite{Hejazi2018,Narula2015,Gazis2015} and websites, such as  \cite{leejH,iotify,lee2018}.

\textbf{Stability}: There are hundreds of platforms in the market, which might have some open issues. Some platforms might fail to deliver services to clients. Thus, a platform should be chosen that has high chances of survival in the market. Information about the platform can be obtained from previous customers using the same platform \cite{Hejazi2018}.

\textbf{Scalability and flexibility}: Initially, a company might be small and operate in a small business area but, ideally, over time, the business will expand and with this growth, the business area will also expand. Thus, to ensure that the IoT platform can support the business throughout its development, the platform should be scalable to business needs \cite{Hejazi2018}. Similarly, the platform should be flexible with regard to the technology, since modern technology and market demands change rapidly. 

\textbf{The pricing model and business case}: Some platform providers offer a low price for a period at the start of a contract agreement, after which the price increases greatly. Additionally, some providers offer a low price to attract customers, but the contract includes limited features and additional features have a significant cost if included.  Thus, a platform should be selected that offers full features for the business at a cost that suits the company’s budget \cite{Hejazi2018},\cite{iiunin2017}.

\textbf{Security}: It is an important aspect of IoT that all platforms  should have with high quality. The security may be a device-to-cloud network security, data encryption, application authentication, secure session initiation, application authentication, cloud security, and device security (authentication and up-to-date certification) \cite{McCle}.

\textbf{Time-to-market}: When selecting an IoT platform, the questions of time-to-market and how the platform provider will support the business during the journey from product conception to sale should be considered. Some IoT platform providers offer quick-start packages for new customers, which can speed up product development, reduce time-to-market and offer better solutions \cite{McCle}.  

\textbf{Data analytics and visualization tools}: Before selecting an IoT platform, prospective IoT platform users should establish which platform offers the best capabilities to aggregate, analyze and visualize data. In particular, users should consider how the IoT platform integrates leading analytics toolsets and uses them to replace built-in functionality. Data analysis and  information visualization requirements should be identified before selecting an IoT platform \cite{Gigli2011}. 

\textbf{Data ownership}: A complicated issue with IoT data is ownership of the data. Different jurisdictions have different laws and legal interpretations. For example, the European union (EU) has different rules and regulations regarding data ownership than the United States (US) \cite{Khakurel}. Therefore, it is important to have knowledge of data rights and the territorial scope of data protection for the IoT platform provider.

\textbf{Ownership of cloud infrastructure}: The hardware infrastructure layer is expensive and some smaller IoT platform providers only offer the software layer. Some providers certify their platform on single or multiple leading public cloud providers and mostly run their services on a single leading platform. The compatibility of the broader enterprise cloud  with the IoT platform provider should be checked \cite{Lama2017}.    

\textbf{Extent of legacy architecture}: The connectivity in an existing IoT is often unknown, and IoT devices are designed to work with a variety of infrastructure systems. Therefore, when selecting an IoT platform, businesses and organizations should ascertain how new generations of technology can interlock with older technology \cite{Matt2016}. 
 
\textbf{Protocol}: The important protocols supported by IoT platforms are MQTT, HTTP, AMQP, and CoAP. Due to its binary nature, MQTT is extremely lightweight and has much lower overheads. As a result of development in technology, new devices are coming onto the market. The selected IoT platform should support new protocols and enable easy upgrade of these protocols \cite{iotify}, \cite{Salman2015}.   

\textbf{System performance}: In an IoT platform, when an event happens, a rule-based trigger might be invoked automatically. Since they support such a method, as larger numbers of devices connect to the IoT platform the average time to analyze and handle each event increases. Prior to the selection of an IoT platform, it should be noted what steps the provider has taken to maintain  IoT platform performance high enough \cite{Vandikas2014}.

\textbf{Interoperability}: The IoT platform solution is a middle-ware. The data collected will be used by many applications and may not be available on the platform  itself. Consequently, the selected IoT platform should support integration with open source ecosystems. Interoperability will enable the organization to gain higher productivity \cite{Koo2017},\cite{Xiao2014}. 

\textbf{Redundancy and disaster recovery}: Problems sometimes occur in the IT infrastructure, either natural or man-made, and IoT platform providers should have dedicated infrastructure to handle data during such occurrences. Issues that require consideration include the data backup plan schedule and whether the IoT platform has failover cluster provision \cite{iotify}.

\textbf{Attractive interface}: The interface provided by the IoT platform should be simple, attractive and user friendly, so that it is easy for customers to use its functionalities. All the services offered to the customers should be easy to access. 

\textbf{Application environment}: Three aspects of the application environment should be considered before selecting an IoT platform: which applications are available out of the box, what are the characteristics of the application development environment, and what are the common interfaces \cite{Lama2017}.

\begin{table*}[t]
\caption{Basic Features provided by the five IoT platforms }
\begin{tabular}{l|c|c|c|c|c}

\multicolumn{1}{l|}{\textbf{Features}}                                                                & \multicolumn{1}{c|}{\textbf{AWS}}                                                                                            & \multicolumn{1}{c|}{\textbf{Microsoft Azure}}                                                       & \multicolumn{1}{c|}{\textbf{Google Cloud IoT}}                                                & \multicolumn{1}{c|}{\textbf{IBM Watson IoT}}                                                                                                   & \multicolumn{1}{c}{\textbf{Oracle IoT}}                                               \\ \hline \hline
\textbf{Security }                                                                                     & \begin{tabular}[c]{@{}c@{}}Link Encryption (TLS),  \\Authentication\\ (Sig V4, X.509)\end{tabular}                    & \begin{tabular}[c]{@{}c@{}}Link Encryption\\  (SSL/TSL)\end{tabular}                    & \multicolumn{1}{c|}{SSL/TLS}                                                         & \begin{tabular}[l]{@{}c@{}}Link Encryption (TLS), \\ Authentication (IBM cloud SSO), \\Identity Management (LDAP)\end{tabular} & REST API                                                                      \\ \hline
\textbf{\revi{Data analytics}}                                                                             & \begin{tabular}[c]{@{}c@{}}Real Time analytics\\ (Rule engine,  Kinesis, \\ AWS Lambda)\end{tabular}    & \begin{tabular}[c]{@{}c@{}}Real Time  \\ analytics\end{tabular}                          & \begin{tabular}[c]{@{}c@{}}Real Time analytics\\ (Cloud IoT Core)\end{tabular}       & \begin{tabular}[c]{@{}c@{}}Real Time analytics \\ (IBM IoT Real time insights)\end{tabular}                                        & \begin{tabular}[c]{@{}c@{}}Real Time \\ \\ analytics\end{tabular}             \\ \hline
\multicolumn{1}{l|}{\begin{tabular}[c]{@{}c@{}}\textbf{\revi{Protocols}}\end{tabular}} & MQTT, HTTP1.1                                                                                                       & \begin{tabular}[c]{@{}c@{}}MQTT, HTTP,\\ AMQP\end{tabular}                                 & MQTT                                                                                 & MQTT, HTTPS                                                                                                                           & MQTT, HTTP                                                                    \\ \hline
\textbf{Visualization tool }                                                                           & AWS IoT dashboard                                                                                                   & web portal                                                                                 & \begin{tabular}[c]{@{}c@{}}Google data studio\\  (Dashboard)\end{tabular}            & web portal                                                                                                                            & web portal                                                                    \\ \hline
\textbf{Data format }                                                                                  & JSON                                                                                                                & JSON                                                                                       & JSON                                                                                 & JSON, CSV                                                                                                                             & \begin{tabular}[c]{@{}c@{}}CSV,\\  REST API\end{tabular}                      \\ \hline
\textbf{\revi{Application Environment} }                                                                        & \begin{tabular}[c]{@{}c@{}}Java, C, NodeJs, Javascript,\\ Python, SDK for Arduino,\\      iOS, Android\end{tabular} & \begin{tabular}[c]{@{}c@{}}.Net, UWP, Jave,\\  C, NodeJS, Ruby,\\ Android,iOS\end{tabular} & \begin{tabular}[c]{@{}c@{}}Go,Java, Python,\\ .NET, NodeJS,\\ php, Ruby\end{tabular} & \begin{tabular}[c]{@{}c@{}}C\#, C, Python,\\  Java, NodeJS\end{tabular}                                                               & \begin{tabular}[c]{@{}c@{}}Java, iOS,\\ Javascript, C,\\ Android\end{tabular} \\ \hline

\end{tabular}
\label{tab-fea}
\vspace{-3ex}
\end{table*}

\textbf{Hybrid cloud}: Some IoT platforms can fit with existing 
IT systems hosted on company premises. In such situations, a hybrid cloud is very useful as mission critical or business sensitive processes can be handled locally, while public and less critical operations can be managed by the  platform \cite{iotify}.   

\textbf{Platform migration}: Over time, and as the company grows, the IoT platform may be unable to meet all the company’s requirements. Thus, a bigger IoT platform provider may be needed. Consequently, companies should ensure that the selected IoT platform provider provides clearly documented interfaces, schema, and API for any possible future migration to other IoT platforms \cite{Hejazi2018},\cite{iotify}. 

\textbf{Previous experience}: Prior to selection, a company should check whether the IoT platform provider has some experience of work similar to that of the company application. Successful working experience in the same area can be considered a good sign \cite{McCle}.

\textbf{Bandwidth}: For efficient movement of information and communication between the processing components, the IoT platform needs low latency and high bandwidth networking. Thus, it should be ascertained that a potential IoT platform provider has a large data pipe and that there is sufficient room to grow \cite{Al-Zihad2018}, \cite{iotify}. 

\textbf{Edge intelligence and control}:  The future of IoT platforms is moving towards distributed, offline and edge intelligence [24]. Devices become more powerful when they are able to make decisions based on local data instead of waiting for every decision from the cloud. Thus, it should be ensured that the IoT platform has the capacity to support new topologies and utilize edge intelligence \cite{Loghin2017}.

\vspace{-1ex}
\section{Research method and data collection}

%
%
%
%

The data collection work began with the literature survey of the research articles and publications related to IoT and its platforms published in different journals, conferences, books. About 200 articles were searched for the selected topic and forty-six were selected for the study. The selected articles are searched in the IEEE, ACM, and Scopus databases, Google Scholar and some websites are searched and referenced. Twenty-one key factors of an IoT platform were identified from the literature.  The preliminary analysis has been published by the authors in a conference paper \cite{Ullah2019}. Relevant parts are summarised again here for clarity within the rest of the paper.
The collected key factors of an IoT platform were verified and categorized using the Delphi method, 
which is an interactive process to collect and distill the data using the judgments of experts using a feedback loop. This method is a flexible research technique that can be used to successfully explore new concepts inside and outside the information system body of knowledge \cite{J.Skulmoski2007}.

\begin{figure}[t]
\centering
\includegraphics[width=0.5\columnwidth]{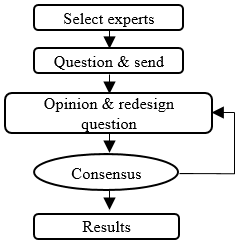}
\caption{Schematic of the Delphi method for verification and categorization.}
\label{Delphi method}
\vspace{-3ex}
\end{figure}

\begin{figure}[t]

  \includegraphics[width=\columnwidth]{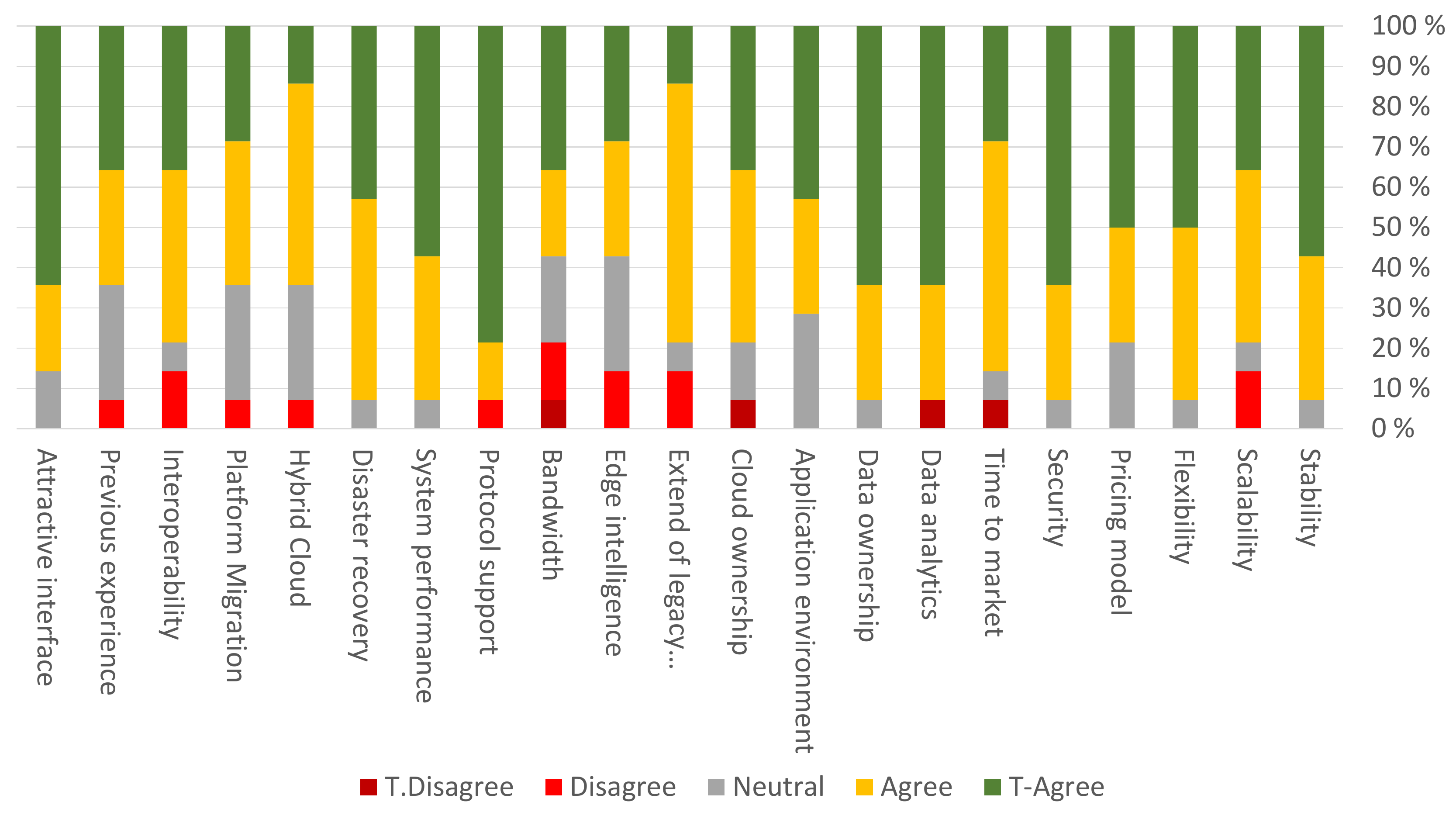}
  \caption{Experts opinion in the first round.}
  \label{Expert opinion 1}
  \vspace{1ex}
  \includegraphics[width=\columnwidth]{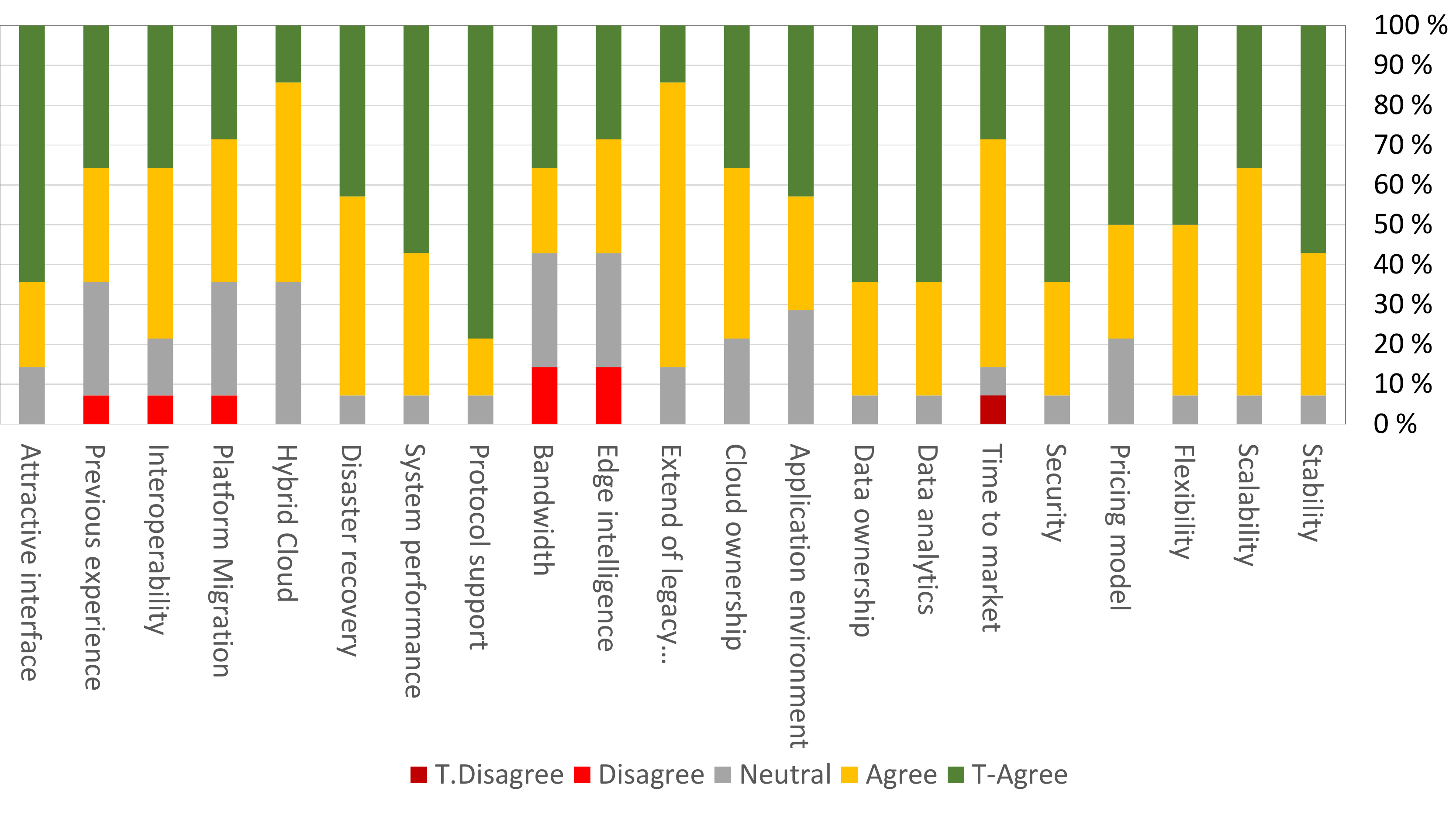}
  \caption{Experts opinion in the second round.}
  \label{Expert opinion 2}
\end{figure}

We employed here a two-round Delphi study, as shown in Fig. \ref{Delphi method}, during the first round of the Delphi study fifteen experts from three different universities were selected based on their experience in IoT field. A questionnaire was designed based on twenty-one questions related to the key factors of IoT platforms \revi{as show in Table \ref{tab-delphi-survey} at Appendix}. A 5-point Likert rating scale was used: (1) totally disagree, (2) disagree, (3) neutral, (4) agree, and (5) totally agree.  The questionnaire was sent to the experts by email to be answered within two weeks. Fourteen experts replied and the response percentage was 93\%. The experts’ opinion of the first round is shown in Fig. \ref{Expert opinion 1}. In the first round, agreed percentage is 80\%, disagree percentage is 6\% and neutral percentage is 14\%.

\revi{There was little conflict between the opinions of the experts about the first round questions as shown in Fig. \ref{Expert opinion 1}, the second round questionnaire was designed based on the experts' opinion of the first round. The original questions were the same as before only the summary of the experts opinion of the first round was subsequently sent to the same experts.} In the second round, fourteen experts replied and the response percentage was 93\%. \revi{In the second round, some of the experts have changed their opinion based on the summary of opinions of first round.} The result is shown in Fig. \ref{Expert opinion 2}. The agreed percentage was then 81\%, disagree percentage is 4\% and neutral percentage is 15\%.
The results of both the rounds are shown in Table \ref{tab-delphi}. Note that for simplicity, we have merged "totally agree" and "agree" to "Agree", and "totally disagree" and "disagree" values to "Disagree". 

 The importance of all the twenty-one factors of an IoT platform were categorized into three categories in the light of experts’ opinions. Factors with agree percentage up-to 79\% and above are considered very important, factors with agree percentage between 78\% and 64\% are considered somewhat important, and factors with agree percentage less than 60\% are considered as less important. According to experts opinions, the factors stability, security, protocol support, system performance, disaster recovery, data analytics, scalability, flexibility, data ownership, extend of legacy architecture, pricing model, interoperability, attractive interface, cloud ownership, and time to market were considered as the most important factors. Four factors, application environment, hybrid cloud, platform migration and previous experience were considered as somehow important and two factors edge intelligence and bandwidth were considered as less important.

\begin{table*}[]
\caption{Results of Delphi study both rounds. The mean and median are taken from the agreed values.}
\begin{tabular}{l|l|c|c|c|c|c|c|c|c|c|c}
                       \multicolumn{2}{c}{}         & \multicolumn{5}{|c|}{\textbf{Survey round 1}}                                                                                                                          & \multicolumn{5}{c}{\textbf{Survey round 2}}                                                                                                                          \\ \hline\hline 
\multicolumn{1}{l|}{\textbf{F}} & \multicolumn{1}{l|}{\textbf{Factor}}   & \multicolumn{1}{l|}{Mean} & \multicolumn{1}{l|}{Median} & \multicolumn{1}{l|}{Disagree \%} & \multicolumn{1}{l|}{Neutral \%} & \multicolumn{1}{l|}{Agree \%} & \multicolumn{1}{l|}{Mean} & \multicolumn{1}{l|}{Median} & \multicolumn{1}{l|}{Disagree \%} & \multicolumn{1}{l|}{Neutral \%} & \multicolumn{1}{l}{Agree \%} \\ \cline{1-12} 
F1  & Scalability   & 4  & 4  & 14\%  & 7\%  & 79\%  & 4    & 4     & 0\%    & 7\%        & 93\%                          \\\hline
F2  & Flexibility   & 4  & 4.5  & 0\%   & 7\%    & 93\%  & 4     & 4.5    & 0\%   & 7\%   & 93\%                          \\\hline
F3                    & Data anlytics                 & 4                         & 5                           & 7\%                              & 0\%                             & 93\%                          & 5                         & 5                           & 0\%                              & 7\%                             & 93\%                          \\\hline
F4                    & Disaster recovery             & 4                         & 4                           & 0\%                              & 7\%                             & 93\%                          & 4                         & 4                           & 0\%                              & 7\%                             & 93\%                          \\\hline
F5                    & Stability                     & 5                         & 5                           & 0\%                              & 7\%                             & 93\%                          & 5                         & 5                           & 0\%                              & 7\%                             & 93\%                          \\\hline
F6                    & Security                      & 5                         & 5                           & 0\%                              & 7\%                             & 93\%                          & 5                         & 5                           & 0\%                              & 7\%                             & 93\%                          \\\hline
F7                    & Data ownership                & 5                         & 5                           & 0\%                              & 7\%                             & 93\%                          & 5                         & 5                           & 0\%                              & 7\%                             & 93\%                          \\\hline
F8                    & Protocol support              & 5                         & 5                           & 7\%                              & 0\%                             & 93\%                          & 5                         & 5                           & 0\%                              & 7\%                             & 93\%                          \\\hline
F9                    & System performance            & 5                         & 5                           & 0\%                              & 7\%                             & 93\%                          & 5                         & 5                           & 0\%                              & 7\%                             & 93\%                          \\\hline
F10                   & Time to market                & 4                         & 4                           & 7\%                              & 7\%                             & 86\%                          & 4                         & 4                           & 7\%                              & 7\%                             & 86\%                          \\\hline
F11                   & Legacy architecture & 4                         & 4                           & 14\%                             & 7\%                             & 79\%                          & 4                         & 4                           & 0\%                              & 14\%                            & 86\%                          \\\hline
F12                   & Attractive interface          & 5                         & 5                           & 0\%                              & 14\%                            & 86\%                          & 5                         & 5                           & 0\%                              & 14\%                            & 86\%                          \\\hline
F13                   & Pricing model                 & 4                         & 4.5                         & 0\%                              & 21\%                            & 79\%                          & 4                         & 4.5                         & 0\%                              & 21\%                            & 79\%                          \\\hline
F14                   & Cloud ownership               & 4                         & 4                           & 7\%                              & 14\%                            & 79\%                          & 4                         & 4                           & 0\%                              & 21\%                            & 79\%                          \\\hline
F15                   & Interoperability              & 4                         & 4                           & 14\%                             & 7\%                             & 79\%                          & 4                         & 4                           & 7\%                              & 14\%                            & 79\%                          \\\hline
F16                   & App. environment       & 4                         & 4                           & 0\%                              & 29\%                            & 71\%                          & 4                         & 4                           & 0\%                              & 29\%                            & 71\%                          \\\hline
F17                   & Hybrid cloud                  & 4                         & 4                           & 7\%                              & 29\%                            & 64\%                          & 4                         & 4                           & 0\%                              & 36\%                            & 64\%                          \\\hline
F18                   & Platform migration            & 4                         & 4                           & 7\%                              & 29\%                            & 64\%                          & 4                         & 4                           & 7\%                              & 29\%                            & 64\%                          \\\hline
F19                   & Previous experience           & 4                         & 4                           & 7\%                              & 29\%                            & 64\%                          & 4                         & 4                           & 7\%                              & 29\%                            & 64\%                          \\\hline
F20                   & Edge intelligence             & 4                         & 4                           & 14\%                             & 29\%                            & 57\%                          & 4                         & 4                           & 14\%                             & 29\%                            & 57\%                          \\\hline
F21                   & Bandwidth                     & 4                         & 4                           & 21\%                             & 21\%                            & 57\%                          & 4                         & 4                           & 14\%                             & 29\%                            & 57\%                          \\\hline\hline
                  --    & \textbf{Percentage}           &     --                      &   --                          & \textbf{6\%}                     & \textbf{14\%}                   & \textbf{80\%}                 & --                & --                   & \textbf{4\%}                     & \textbf{15\%}                   & \textbf{81\%}      \\\hline  
\end{tabular}
\label{tab-delphi}
\vspace{2ex}
\end{table*}

\section{Proposed method to compare IoT Platforms}

\subsection {\revi{Identifying IoT platforms}}
\vspace{-1ex}
IoT applications need a platform to run smoothly and handle the data so that companies can take future decisions based on the data processed by the IoT platform \cite{dani2017}. Hundreds of IoT platforms are available and finding the most suitable IoT platform for a specific IoT application is becoming increasingly difficult. The problem is compounded by a lack of experience and knowledge, and in some cases, a company may select a platform without adequate requirements analysis, which later leads to problems \cite{Hejazi2018}. When developing an IoT application for business needs, IoT platforms are the first place that can provide the facilities for deploying and running the business application \cite{Nakhuva2015}. There is high competition between the various IoT platforms in the market. In this research, we have arbitrarily selected five well-known platforms based on information collected by specialized websites\footnote{For example: \url{https://internetofthingswiki.com/top-20-iot-platforms/634/}} and reports about their market share (e.g., \cite{Reno2018}), which we will then use to demonstrate our methodology to support selecting appropriate platform solutions. The selected IoT platforms are: Amazon Web Services (AWS) IoT,  Google Cloud IoT, Microsoft Azure IoT suite, IBM Watson IoT, and Oracle IoT. Their basic technical features are compared in terms of security, \revi{ data analytics, protocols, visualization tool, data format,  and application environment}, as  shown in Table \ref{tab-fea}. Next we will briefly present these platforms based on their own descriptions and other specialized references, trying to mimic how organizations collect information for selecting the service provider.

\begin{table*}[!ht]
\centering
\caption{Factors for selecting IoT platform factors}
\begin{tabular}{l|l|l}
Source & Year & Factors                                            \\\hline\hline
\cite{Agarwal2018}      & 2018 & stability, time to market, pricing model, protocols, data analytics     \\\hline
\cite{Hejazi2018}       & 2018 & stability, scalability, flexibility, pricing model                                        \\\hline
\cite{Narula2015}       & 2015  & Security                                     \\\hline
\cite{Gazis2015}        & 2015 & interoperability \\\hline
\cite{kondra2018}       & 2018  & interoperability, security, reliability, protocols, data analytics, interactive interface. \\ \hline     
\cite{Ganguly2016}      &2016  &protocols, data analytics, scalability, security.\\\hline
\cite{victor2019}       &2019  &interactive interface,interoperability,security,connectivity services,device management.\\\hline
\cite{maker2018}        &2018  &protocol, security, bandwidth, application environment, cost.\\\hline
\cite{Ismail2019}       &2019  &scalability, stability.\\\hline
\end{tabular}

\label{tab-prev}
\end{table*}

\textit{Amazon Web Services (AWS):} AWS was launched in 2006 and is the leading platform with 33\% market share in 2018 \cite{Reno2018}. AWS  provides storage space, compute capability, data management and other infrastructure resources \cite{Nakhuva2015}. It also offers artificial intelligence (AI) services \cite{amazon2019}. AWS has customers like Dropbox, Netflix, and Philips \cite{Rojas2017}.

\textit{Microsoft Azure:} Azure was launched in 2010 and had in 2018 a market share of 24\% \cite{Reno2018}. It is capable of data gathering, processing, storing and using analytics.
It also allows IoT applications to work in two-way communication \cite{Azure2019}. Azure has customers like Apple-iCloud, EasyJet, and Xerox.

\textit{Google Cloud IoT (GCP):} GCP was launched in 2008 having market shares of 12\% in 2018 \cite{Reno2018}. GCP uses cloud and edge computing. It offers data analytics and machine learning while employing Google Maps to  track the assets' positions.  GCP has customers like PayPal and Bloomberg. \cite{google2019,Challita2018}.

\textit{IBM Watson IoT platform:} IBM Watson had a market share of 18\% in 2018 \cite{Reno2018}  and provides  connectivity, analysis, device management and information management \cite{Salami2018,IBMwat2019}. IBM Watson employs two-way communication with the end user and also uses blockchain services. The main customers are STAPLES and AUTODESK.

\textit{Oracle IoT platform:} Oracle\footnote{Its market share is not available in \cite{Reno2018}; our selection was based on \url{https://www.zdnet.com/article/top-cloud-providers-2018-how-aws-microsoft-google-ibm-oracle-alibaba-stack-up/}. }
offers acquisition, analysis and integration of data \cite{Ray2017}, also using edge analytics  \cite{things2019}. The main customers of Oracle IoT platform are Softbang LLC and Anson McCade.

\subsection {\revi{Proposed method}}

Previous studies have also sought to identify relevant factors for selecting an IoT platform for business. Table \ref{tab-prev} summarises these. Whilst there is considerable overlap none of these prior studies have identified the number or granularity of factors as in our approach. Therefore, while they are no doubt suitable for the specific applications and domains within which they were developed, we have aimed to create a more general approach that can be more widely used across all these cases.
To show how our general framework can be applied to assessing and choosing an IoT platforms, in this study we have selected the top five IoT platforms based on market share.
We have compared these IoT platforms according to the twenty-one key IoT platform factors that we have identified from the literature and verified using Delphi study. 
We have compared these twenty-one key factors with the features provided by those selected five IoT platforms as shown in Table \ref{tab-feat}.

More specifically, the entries of Table \ref{tab-feat} have the following meaning related to the specific feature to be considered: `yes' means the feature is available, `high' indicates strong, `bad' shows weak, `good' indicates that the feature is very good, `-' shows that the feature is unknown and `no' indicates that the feature is not available in the platform. In order to identify and fill the features of the selected five IoT platforms, different articles \cite{Ray2017,Agarwal2018,Hejazi2018,Ammar,Challita2018,Ullah2019,Nakhuva2015} have been studied from many databases. Some websites \cite{amazon2019,google2019,IBMwat2019,Azure2019,things2019,iotify} have been used, especially the websites of those selected IoT platforms. A few white papers \cite{Cloud2018} have also been studied.

\begin{table*}[!ht]
\caption{Reflecting the twenty-one key IoT platform features in the five main IoT platforms }
\centering
\begin{tabular}{l|c|c|c|c|c}
\textbf{Factors}         &\textbf{AWS} & \textbf{Azure} & \textbf{Google cloud} & \textbf{IBM Watson} & \textbf{Oracle IoT }\\\hline \hline
Scalability                   &yes     &yes       &yes            & yes        & yes           \\\hline
Flexibility                   &yes     &-         &yes            &-            & yes           \\\hline
Data analytic                 &yes     &yes       &yes            &yes         & yes           \\\hline
Disaster recovery             &yes     &yes       &no             &no          &no            \\\hline
Stability                     &yes     &yes       &yes            &-           &-           \\\hline
Security                      &high    &high      &high           &high        &high           \\\hline
Data ownership                &-        &yes       &-               &-            &-            \\\hline
Protocol support              &yes     & yes      &-               &yes         & yes           \\\hline
System performance            &yes     &-          &yes            &yes         &-            \\\hline
Time to market                &yes     &yes       &-               &-            & yes           \\\hline
legacy architecture           &yes     &-          &-               &-            & yes           \\\hline
Attractive interface          &yes     &yes       &-               &no          &-            \\\hline
Pricing model                 &bad     &bad       &good           &-            &-            \\\hline
Cloud ownership                &yes    &yes       &yes            &-            &yes            \\\hline
Interoperability              &yes     &-          &-               &-            &yes            \\\hline
App. environment              &yes     &yes       &yes            &yes         &yes            \\\hline
Hybrid cloud                  &yes     & yes      &-               &-            &-            \\\hline
Platform migration            &yes     & yes      &-               &-            &-            \\\hline
Previous experience           &yes     &yes       &-               &-            &-            \\\hline
Edge intelligence             &yes     &yes       &yes            &-            &yes            \\\hline
Bandwidth                     &-        &-          &good           &-            &-           \\\hline
\end{tabular}
\label{tab-feat}
\vspace{2ex}
\end{table*}


\begin{figure*}[!ht]
  \begin{center}
  \includegraphics[width=1.75\columnwidth]{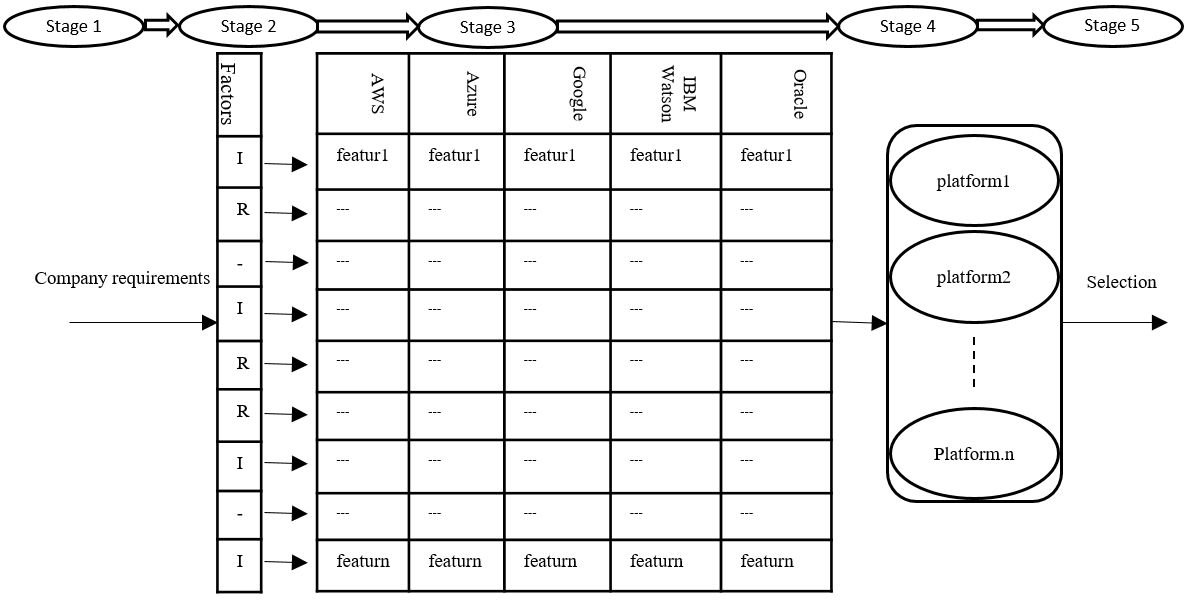}\\
  \label{Key features}
  \caption{Comparing Key factors with the features offered by the IoT platform.}\label{Key features}
  \end{center}
  \vspace{-0.6cm}
\end{figure*}

The framework for selection of an IoT platform is illustrated in Fig. \ref{Key features} as a schematic of the selection procedure. The whole process consists of five stages. In the first stage the company finalize their business requirements. In the second stage the company requirements are applied to prioritising which factors are required (R), important (I) and not required (-) for this business context. In the third stage the R and I factors are compared with the features provided by the five selected IoT platforms. The IoT platform/s that provide a maximum of the features as compared to the requirements are selected and shifted to the stage four. In stage four there might be one or many IoT platforms that match the required and important factors. Stage five is the decision, which is explained next.

If there is one IoT platform that provides the most required and important features then the same IoT platform can be selected for the business application. But, if there are multiple IoT platforms providing these features then the company may choose an IoT platform based on the comparison of their match to I factors like pricing, time to market etc. and select a suitable IoT platform for their business needs. There might also be chances that none of the platforms provide all of the required features; this might indicate that new platforms should be selected and evaluated accordingly.

The five stages of the framework are explained in a simple example. A selling company is interested to be on the web and use IoT application for its business. \revi{ Initially they were interested to learn the components of IoT to understand what is IoT and how it works. Secondly they were interested to know what is an IoT platform and what are key factors of an IoT platform}. They need an IoT platform for their business application but they do not know which \revi{platforms are providing what features and which } one is best.\revi {when the company have the knowledge of IoT, IoT platform factors and the features those platforms are providing} then in stage 1, the company go through each of the 21 factors that have been identified as important in choosing a platform and use this to help them to formulate their business requirements. In stage 2, these factors are prioritized as being either required (R), important (I),  or not required (-) for their business needs. They find that their required factors to consider (R) are scalability, time to market, flexibility. Their important factors to consider (I) are pricing and interoperability. In stage 3, the R and I factors are compared with identified features of IoT platforms. AWS and Oracle are the platforms that are known to match all required features; both have the feature of interoperability, but AWS has worse pricing model while one for Oracle is unknown. The company may request the pricing model from Oracle  and then choose based on this.

\vspace{-0.2cm}
\section{Concluding remarks}

The aim of this study is to build an objective methodology to support  organizations to select the most suitable IoT platform based on their specific needs. To do so, we first reviewed the building blocks of IoT explaining how they are combined to perform specific tasks.  Second, we identified twenty-one key factors of IoT platforms from the literature and then verified with expert’s opinion using Delphi studies.  Finally, we have designed a theoretical framework for selection of IoT platform and tested it in five well-known examples. This research then provides a general framework to select the most suitable IoT platform for a specific organization by comparing its specific requirements with the features offered by the different platforms. As future work, we expect to evaluate the IoT platforms in different vertical cases like energy and industry 4.0. Our goal is to build an automated procedure that also includes the possibility of weighting the factors based on interviews with experts, developers and programmers from that particular domain.

\vspace{-0.2cm}
\appendix
\vspace{-0.2cm}
Table \ref{tab-delphi-survey} shows the questions used to carry out the Delphi method employed to validate the proposed 21 key factors.

\begin{table*}[t]
\centering
\caption{Questions used in survey, during the Delphi method}
\begin{tabular}{l|l}
\textbf{Q\#} & \textbf{Survey question}\\ \hline\hline
Q1 & What is your opinion about the importance of stability of IoT platform?\\ \hline
%
Q2 & What is your opinion about the importance of Scalability of the enterprise of IoT platform? \\ \hline
Q3 & Do you think that IoT platform should be flexible with the advancement of technologies?\\ \hline
Q4 & Do you think it is important to know about the pricing models before selecting IoT platform?\\ \hline
Q5 & Do you think IoT platform should provide security at both the ends, software and hardware? \\ \hline
Q6 & Do you think IoT platform can reduce Time to market for the business?\\ \hline
Q7 & Do you think IoT platform should support the basic descriptive, predictive and perspective analytics?\\ \hline
Q8 &  Do you think it is important to know who will own the data collected by IoT platform?\\ \hline
Q9 & Is it important to know the application environment of IoT platform?\\ \hline
Q10 & Do you think it is important to know the Ownership of cloud infrastructure?\\ \hline
Q11 & Do you think extend of legacy architecture in IoT platform is important?  \\ \hline
Q12 & Do you think Edge intelligence is important for IoT platform?\\ \hline
Q13 & Do you think IoT platform needs high bandwidth networking? \\ \hline
Q14 & Do you think it is important for IoT platform to support new Protocols and its updated versions? \\ \hline
Q15 &Do you think the IoT platform venders should implement some steps to keep System performance high?\\ \hline
Q16 & Do you think the IoT platform providers should have some dedicated infrastructure to handle customer data if there is some problem in IT infra? \\ \hline
Q17 & Do you think Hybrid cloud is important for IoT platforms?\\ \hline
Q18 & Do you think IoT platform providers should provide facilities to customers for any possible migration to other IoT platform in future?\\ \hline
Q19 &Do you think IoT platform Interoperability will enable the organization to get higher productivity?\\ \hline
%
Q20 & Is it necessary to check the previous experience of IoT platform, before selection? \\ \hline
%
Q21 & Is it necessary that user interface of the IoT Platform should be simple and attractive?\\ \hline
\end{tabular}
\label{tab-delphi-survey}
\vspace{2ex}
\end{table*}

\vspace{-2ex}
\bibliographystyle{IEEEtran}
\bibliography{photo/Bibliography}

\end{document}